\documentclass[fleqn,usenatbib]{mnras}

\usepackage{mathptmx}

\usepackage[T1]{fontenc}

\DeclareRobustCommand{\VAN}[3]{#2}
\let\VANthebibliography\thebibliography
\def\thebibliography{\DeclareRobustCommand{\VAN}[3]{##3}\VANthebibliography}

\usepackage{graphicx}	
\usepackage{amsmath}	
\usepackage{amssymb}	

\usepackage{hyperref}
\usepackage{parskip}
\usepackage{mathrsfs}
\usepackage{makecell}

\usepackage{soul}

\newcommand{\fml}{\texttt{FML}}
\newcommand{\mtm}{\texttt{map2map}}
\newcommand{\LCDM}{$\Lambda$CDM}
\newcommand{\cola}{\texttt{COLA}}
\newcommand{\fRnot}{$|f_{R_0}|$}

\title[A field-level emulator for modified gravity]{A field-level emulator for modified gravity}

\author[D. Saadeh et al.]{
Daniela Saadeh,$^{1}$\thanks{E-mail: daniela.saadeh@port.ac.uk}
Kazuya Koyama,$^{1}$
Xan Morice-Atkinson$^{1}$
\\
$^{1}$Institute of Cosmology \& Gravitation, University of Portsmouth, Dennis Sciama Building,
Burnaby Road, Portsmouth, PO1 3FX, United Kingdom
}

\date{Accepted XXX. Received YYY; in original form ZZZ}

\begin{document}
\label{firstpage}
\pagerange{\pageref{firstpage}--\pageref{lastpage}}
\maketitle

\begin{abstract}   
Stage IV surveys like LSST and Euclid present a unique opportunity to shed light on the nature of dark energy. However, their full constraining power cannot be unlocked unless accurate predictions are available at all observable scales. Currently, only the linear regime is well understood in models beyond $\Lambda$CDM: on the nonlinear scales, expensive numerical simulations become necessary, whose direct use is impractical in the analyses of large datasets. Recently, machine learning techniques have shown the potential to break this impasse: by training emulators, we can predict complex data fields in a fraction of the time it takes to produce them.

In this work, we present a field-level emulator capable of turning a $\Lambda$CDM N-body simulation into one evolved under $f(R)$ gravity. To achieve this, we build on the \texttt{map2map} neural network, using the strength of modified gravity \fRnot\ as style parameter. We find that our emulator correctly estimates the changes it needs to apply to the positions and velocities of the input N-body particles to produce the target simulation.

We test the performance of our network against several summary statistics, finding $1\%$ agreement in the power spectrum up to $k \sim 1$ $h/$Mpc, and $1.5\%$ agreement against the independent boost emulator \texttt{eMantis}. Although the algorithm is trained on fixed cosmological parameters, we find it can extrapolate to models it was not trained on. Coupled with available field-level emulators and simulation suites for $\Lambda$CDM, our algorithm can be used to constrain modified gravity in the large-scale structure using full information available at the field level.\\
\end{abstract}




\section{Introduction}

Precise data from numerous cosmological probes conclusively show that the expansion of the Universe started accelerating in the recent past. Within general relativity, our standard theory of gravity, this observation can only be accounted for with the introduction of an exotic constituent -- dark energy -- behaving like a fluid with negative pressure, and making up approximately $\sim70\%$ of the energy budget today (\citet{Riess_1998,Perlmutter_1999}; also see \citet{Planck,DES2024,DESI2024}. Another possibility is that we need to modify our theory of gravity at cosmological scales, where it is currently poorly constrained.

There is some overlap between these two scenarios \citep{Joyce_2016}: a dark energy model introduces a new field that may mediate extra interactions with other constituents in the Universe; the growth of structure under the combined effect of standard gravity and the extra interactions can then look as it would under a modified law of gravity. Indeed, the recent Baryon Acoustic Oscillation measurements by Dark Energy Spectroscopic Instrument (DESI), combined with Planck cosmic microwave background and supernovae measurements, hinted that dark energy might be dynamical \citep{DESI2024}. 

Galaxy clustering and weak lensing data contain highly constraining information on the distribution of matter in the Universe, which is highly sensitive to the growth of structure and the agents shaping it, including dark energy. The traditional approach in the analysis of such cosmological probes is to estimate summary statistics (such as the galaxy or matter power spectra) from the data, and compare them against theoretical predictions for the same estimator. Although powerful, this approach misses important information in the form of phases \citep{Soda, COLES1, COLES2, Bacon}, which contain most of the morphological information. Indeed, recent work has shown that including phase information greatly improves the constraints from the large-scale structure \citep{nguyen2024information}.

Recent years have seen considerable effort to move beyond the powerful, but lossy, data compression offered by summary statistics, towards field-level inference -- see for example \cite{BORG,Dvorkin, Lemos, WL, WL2}. Characterising the distribution of matter at scales where the local densities are large is challenging: no analytical solution is available, and sophisticated numerical simulations become necessary. In modified gravity models, we are further required to solve additional equations for the extra fields. To evade stringent local tests of gravity, these models typically invoke screening mechanisms realised by nonlinearities of the additional field \citep{Joyce_2016, koyama}, which make simulations even more expensive and challenging.

 \begin{figure*}
	 \centering
	 \includegraphics[width=\textwidth]{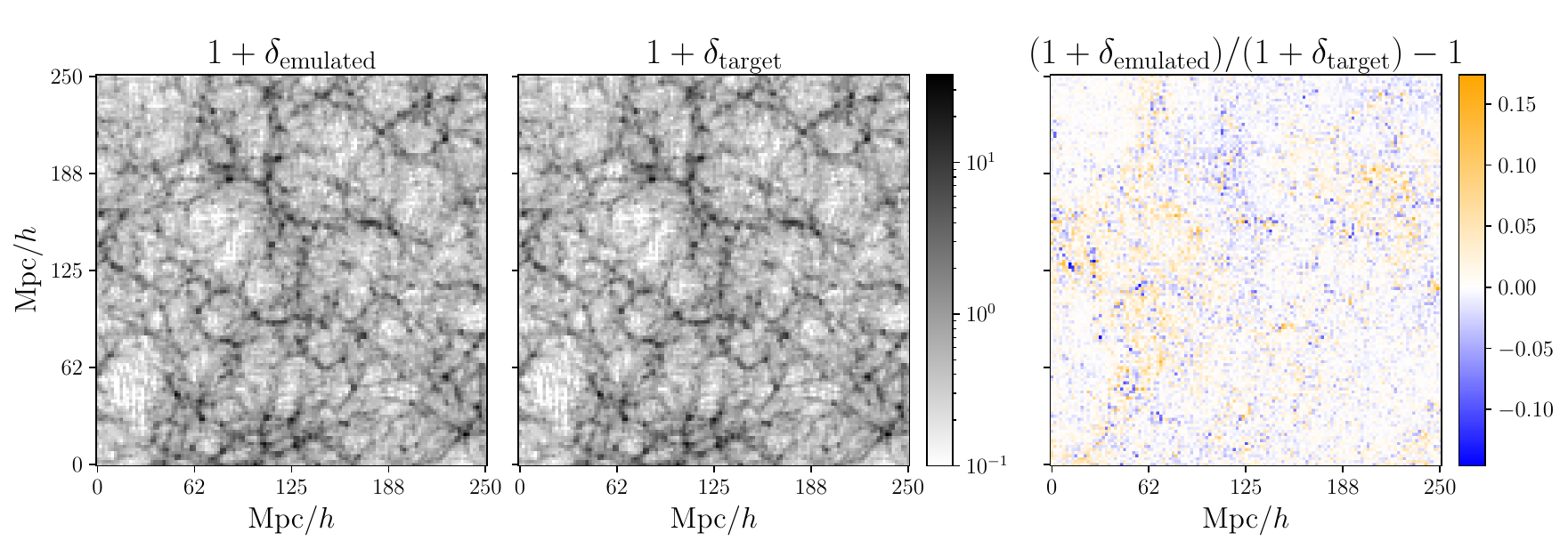}
	\caption{Comparison of the emulator's output against the target snapshot. The plot shows a slice $250$ Mpc$/h$ $\times$ $250$ Mpc$/h$ wide and $29.3$ Mpc$/h$ deep; the strength of modified gravity is \fRnot$=10^{-5}$. The left and middle columns show the emulated and target overdensity, whereas the rightmost column shows the relative error on the emulator's output.}
	\label{fig:m2m_vs_real}
 \end{figure*}

Several numerical codes have been developed for this purpose \citep{Winter2015}, possessing different trade-offs between accuracy and speed. Among the most accurate, but slowest, are the adaptive-mesh-refinement codes, which refine higher density regions in a simulation box until the spatial resolution necessary for the target accuracy is reached. One of these codes is \texttt{RAMSES} \citep{RAMSES}, which was extended to work with modified gravity in \texttt{ECOSMOG} \citep{ECOSMOG}. Other approaches sacrifice some accuracy on the smaller scales in favour of a considerable speed up: this is the case with the particle-mesh code \texttt{MG-GLAM} \citep{MG-GLAM_conformal,MG-GLAM_derivative} and the \fml{} library \citep{FML}, which implements the \cola{} algorithm \citep{COLA, MGCOLA, MGCOLA2, hiCOLA}. In the latter case, the displacement from an initial grid is divided into a large-scale contribution that can be predicted analytically with Lagrangian Perturbation Theory (LPT), and a small-scale term that is determined using particle-mesh N-body force calculations. This approach allows us to capture the large-scale clustering accurately while using considerably less time steps than would otherwise be necessary in integration. For this reason, the \cola{} algorithm is widely used in tasks that require generating a large number of catalogues, for instance in the estimation of the covariance matrix. 

Even with the impressive speed-ups brought about by the \texttt{MG-GLAM} or \cola{} codes, the number of CPU hours necessary to evolve an N-body simulation for a single set of cosmological parameters makes their use in cosmological statistical analyses prohibitive \citep{nbody}. The problem is exacerbated in simulations of modified gravity, because of the extra nonlinear equations to solve for. Consequently, simplified approaches have been used in tests of modified gravity on nonlinear cosmological scales, where analytical linear theory has been extended beyond its regime of validity, or predictions at the nonlinear scales have been modelled by extrapolating general relativity. This approach risks missing unique signatures of modified gravity.

With scientific data soon coming from Euclid \footnote{https://www.euclid-ec.org} and the Vera C. Rubin Observatory \footnote{https://www.lsst.org}, there is urgency in refining our analysis methods so that they are fit for the next generation of surveys: by some estimates, most of the new information coming from the new experiments will be at nonlinear scales \citep{2019A&A...625A..64J}.

Over the last decade, \textit{emulators} were developed, i.e. algorithms capable of reproducing complex data fields starting from simpler inputs. \cite{Angulo_2010} used rescaling techniques to show that a \LCDM\ simulation could be turned into another one with similar, but different, cosmological parameters, if lengths, velocities, masses were appropriately rescaled and redshifts relabelled. Further improvement to larger-scale modes could be obtained by updating the particle positions and velocities at linear scales, so that they followed the behaviour predicted for large scales by the Zel{'}dovich approximation in the cosmology of the target simulation. Subsequently, \cite{Mead_2015} applied rescaling techniques to modified gravity: starting from a parent simulation with a large clustering amplitude -- so that it could mimic the effects of an additional attractive force -- the authors showed they could reproduce several features of a lower-clustering simulation run under $f(R)$ gravity. Considerable improvement was obtained after modifying the large scales in the rescaled simulation to follow the scale-dependent modifications predicted by the Zel{'}dovich approximation; additionally, the accuracy of the redshift space distortions could be considerably improved upon by restructuring haloes to match target concentrations and velocity dispersions. By applying these analytical transformations, the algorithm of \cite{Mead_2015} was able to obtain $5\%$ agreement in the density power spectrum computed from the emulated snapshot, up to scales $k < 1$ $h/$Mpc. Both the works of \cite{Angulo_2010} and \cite{Mead_2015} (as well as follow-ups) were able to achieve remarkable accuracy given a few well-understood analytic prescriptions; on the other hand, they fall short of the necessary accuracy for Stage IV experiments. 

Recent years have seen considerable effort towards using machine-learning methods to create more powerful, and faster, emulators. In this case, the underlying principle is that we can train an algorithm to reproduce a complicated cosmological estimator or field by showing it a large number of examples. Although the amount of time necessary to build the dataset to train and validate such an algorithm is inevitably large, the effort to produce one needs to be made only once, with the algorithm subsequently reproducing that output in a fraction of a second. So far, the main focus has been the emulation of the power spectrum (see for example \cite{EE2, Bacco, COSMOPOWER, matryoshka}, which was recently extended to cover modified gravity models \citep{FORGE, eMantis, Fiorini2023}. Other work \citep{Lombriser_2016,React_1,React_2,React_3} employed an approach based on the halo model to predict the nonlinear matter power spectrum in modified gravity; \citet{React_3} also used a phenomenological description to capture the effects of several screening mechanisms at once.

Other works have focused on field-level inference: \texttt{D$^3$M} \citep{D3M} was able to map from the initial conditions of a quasi-N-body simulation (produced with \texttt{FastPM}, \citet{FastPM}) to its output at redshift $z=0$, using convolutional neural networks. Trained on a fixed set of parameters, the network was found to be able to extrapolate to different cosmologies from the one it was trained on. Subsequently, \texttt{NECOLA} \citep{NECOLA} was able to map from the output of quasi-N-body \cola{} codes to the full-N-body simulations of the \texttt{QUIJOTE} suite \citep{QUIJOTE}. In that work, it was noted that the mapping was largely insensitive to cosmological parameters. Subsequently, \cite{map2map} presented \mtm, an emulator of \LCDM\ simulations using \textit{style} parameters to characterise the output dependence on the matter density parameter $\Omega_M$. For the \LCDM{} model, \mtm\ took snapshots produced under the Zel'dovich approximation as input, and produced full-N-body snapshots as output. The emulator was made up of two convolutional neural networks each producing the non-linear displacement fields and the velocities of N-body particles. The algorithm was found to reproduce successfully the matter power spectra within $\sim 1\%$ and up to $k \sim 1$ $h/$Mpc, also showing good agreement in bispectra, halos and the redshift space distortions. Furthermore, \citet{Jamieson_2022} tested the same machine-learning model on configurations for which the evolution of structure is known, like a spherical distribution or isolated plane waves, finding  it was reproducing known solutions correctly.

In this work, we present a field-level emulator of the large-scale structure in modified gravity, building on \mtm{}. Our approach is to learn how to modify a \LCDM\ simulation so that it becomes one under modified gravity, for some input strength of that modified gravity encoded as style parameter. A large suite of \LCDM\ simulations is already available, for example, in the \texttt{QUIJOTE} simulations \citep{QUIJOTE}: using our emulator, we can turn these into modified gravity simulations, which are considerably more expensive to run.

Learning to map the difference between \LCDM\ and modified gravity makes our emulator less sensitive to the specifics of a particular large-scale structure code, allowing us, potentially, to apply our approach to input snapshots produced under different N-body codes. This approach has been shown to be successful in previous work emulating the matter power spectrum, although the applicability to higher-order statistics remains to be verified. In particular, \citet{Gui} showed that emulating the power spectrum ratio of two different models was more efficient than attempting to reproduce the power spectrum in modified theories of gravity directly. In fact, this choice led to spurious effects due to cosmic variance or resolution largely cancelling out. Our field level emulator extends this approach beyond the power spectrum. Recently, a similar idea has been developed to generate mock halo catalogues rapidly, by using a mapping relationship based on the dark matter density field of a $\Lambda$CDM simulation \citep{Garcia-Farieta:2024xbn}. 

To illustrate the concept, we focus in this work on $f(R)$ gravity, an extensively-studied theory whose behaviour is well understood, and for which independent numerical codes are available to test the robustness of our method. The approach is straightforward to generalise to other theories of gravity.

In order to build our algorithm, we produce training, validation and datasets by using the quasi-N-body code \fml{}: our algorithm can then be used in tandem with \mtm\ to produced arbitrary cosmological simulations starting from \LCDM\ simulations. Following \cite{map2map}, we use a pair of neural networks to learn the Lagrangian displacements and velocities of dark matter particles, from which the density and momentum fields can be then obtained in post-processing.

This paper is organised as follows. In section \ref{sec:Method}, we discuss the mapping that we want the network to learn, the dataset used and the machine learning model we employ.
In section \ref{sec:Results}, we test the accuracy of our emulator against several summary statistics; then we compare the boost factor inferred from the emulated snapshots against \texttt{eMantis}, and the snapshots themselves against the rescaling algorithm of \citet{Mead_2015}. Finally, we test our emulator against models characterised by cosmological parameters it was not trained on, finding that it is able to extrapolate. We conclude in section \ref{sec:Conclusion}.

\section{Methods}
\label{sec:Method}

\subsection{Mapping} \label{sec:mapping}
The purpose of N-body simulations is to evolve the growth of cosmic structure over the age of the Universe. In practice, this is done by tracking the positions and velocities of particles that are initially uniformly distributed in the very early Universe, taken to represent large masses $m$ of dark matter. The $m$ value is taken to be the same for all particles.

In the Lagrangian approach, particles are initially distributed on a uniform grid of positions $\mathbf{q}$. As the system evolves under the action of gravity, their positions at time $t$ are displaced to $\mathbf{x}(\mathbf{q};t) = \mathbf{q} + \boldsymbol{\Psi}(\mathbf{q}; t)$, where we have labelled particles by their initial Lagrangian position. The quantity $\boldsymbol{\Psi}(\mathbf{q};t)$, called the \textit{displacement}, can then be used in place of the original particle position. We can similarly consider the particle velocities $\mathbf{v}(\mathbf{q})$, and also label them by their initial position.

From the particle positions and velocities, one can reconstruct the density and momentum fields. For the former, we can write
\begin{equation}
	\delta( \mathbf{x} )  = \frac{ n( \mathbf{x} )}{\bar{n}} - 1
	\label{eq:Eul_density}
\end{equation}
where $n(\mathbf{x})$ stands to represent the number of particles in the voxel at position $\mathbf{x}$ and $\bar{n}$ is the average number count across the whole box. These are multiple algorithms to assign the mass of particles to voxels, trading off accuracy and speed as they aim to attenuate spurious effects from the discretisation of a continuous field into particles. Here, we use the second-order cloud-in-cell (CIC) scheme, where every particle is treated as a cube of uniform density and one grid-cell wide, therefore contributing its mass to as many as eight neighbouring voxels.

For the momentum field, we can write:
\begin{equation}
	\mathbf{P}( \mathbf{x} ) = m \mathbf{v}( \mathbf{x} )
	\label{eq:Eul_momentum}
\end{equation}
where $\mathbf{v}(\mathbf{x})$ is the velocity field at the voxel defined by position $\mathbf{x}$. Similarly to the density field, we use the CIC to reconstruct the velocity field from the particle positions and velocities, meaning that the velocity of a particle can be deposited to up to eight neighbouring voxels. The final velocity field at a voxel is then the vector sum of the velocity deposits it received. At empty voxels, the momentum field is better defined than the velocity field: in the real Universe, velocities in voids can be large whereas, in simulations, they are ill-defined when no particle is present; the momentum is free from this issue.

In this work, we train the network to map the final \LCDM{} displacements at redshift $z=0$ to the equivalent displacements under $f(R)$ gravity:
\begin{equation}
\label{eq:displacement}
	\boldsymbol{\Psi}^{(\Lambda \mathrm{CDM})}(\mathbf{q};z=0) \mapsto \boldsymbol{\Psi}^{(\mathrm{MG})}(\mathbf{q};z=0).
\end{equation}
For the velocities, we similarly train the network to perform the mapping:
\begin{equation}
\label{eq:velocity}
	\mathbf{v}^{(\Lambda \mathrm{CDM})}(\mathbf{q};z=0) \mapsto \mathbf{v}^{(\mathrm{MG})}(\mathbf{q};z=0).
\end{equation}
The difference from \LCDM\ is the most important quantity to learn correctly, as it is ultimately the discriminator between different theories of gravity.

\subsection{Training, validation, and test sets} \label{sec:datasets}

 \begin{figure}
 	\includegraphics[width=\columnwidth]{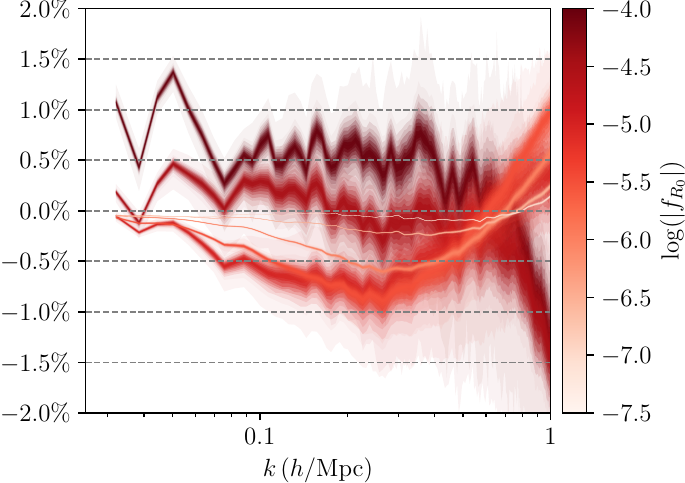}
	\caption{Relative error on the boost factor ${P(k)}^{\mathrm{(MG)}} / {P(k)}^{\mathrm{(\Lambda CDM)}}$ as computed from the 2000 \fml{} simulations in the traning, validation and test datasets, and using the screening efficiencies listed in Table~\ref{tab:se}. For this comparison, we use \texttt{eMantis} as the ground truth. Results for $\log\left|f_{R_0}\right|=-7.5$ are not shown because this value is outside the emulation range of \texttt{eMantis}. However, note that the impact of the screening efficiency is much less pronounced as the model approaches \LCDM\ more closely.}
	\label{fig:se_boxplot}
 \end{figure}

We obtain the simulations in the training, validation and test sets using the publicly-available \fml{}\footnote{\href{https://github.com/HAWinther/FML}{https://github.com/HAWinther/FML}} library \citep{FML, Fiorini:2022srj}. Although the \cola{} method is approximate, it was shown to reproduce the modified gravity boost accurately: in other words, although a single \cola{} simulation may lack power on the small scales compared to a full-N-body simulation, the \textit{difference} between two cosmological models is typically rendered very well \citep{Gui}.

We run simulations of $N_p=512^3$ dark matter particles over a box of $L=1$ Gpc$/h$: these values match the specifications of the \texttt{QUIJOTE} simulations, so that this work can be used in tandem with other emulators trained on \texttt{QUIJOTE}. For all simulations, we use $50$ time steps and, following \citet{Albert}, and use a force grid of $N_g = 3 \times N_p$ nodes. To decrease the overall size of the dataset, we use single-precision simulations: this results in displacement and velocity arrays of $\sim1.6$ Gb each.

For all simulations, we set $\Omega_{\mathrm{CDM}}=0.235$, $\Omega_b=0.046$, $\Omega_{\Lambda}=0.719$, $h = 0.697$, $n_s = 0.971$, $\sigma_8 = 0.842$, where $\Omega_{\mathrm{CDM}}$, $\Omega_b$ and $\Omega_{\Lambda}$ are the dark matter, baryonic and dark energy density, respectively, $h$ is the dimensionless Hubble parameter, i.e. $H_0 = h \times 100$ $\mathrm{km}\, \mathrm{s^{-1}} \mathrm{Mpc^{-1}}$, $n_s$ is the scalar spectral index, and $\sigma_8$ is the power spectrum amplitude at the scale of 8 $\mathrm{Mpc}/h$. We do not include any radiation density or neutrinos. 

In our simulations, we vary the strength of modified gravity, regulated by the parameter \fRnot, within the discrete values $\log_{10} |f_{R_0}| \in S_{|f_{R_0}|} \equiv \{ -7.5, -7, -6.5, -6, -5.5, -5, -4.5, -4 \}$. With this choice, general relativity corresponds to \fRnot$=0$. We do not vary other cosmological parameters, which we defer to future work: however, we do assess the performance of our network in extrapolation in Sec.~\ref{Sec:extrapolations}, for the models reported in Table~\ref{tab:extrapolations}. Note that the ratio between the power spectrum in $f(R)$ and \LCDM{}, highly sensitive to modified gravity, is largely insensitive to cosmological parameters \citep{Winter2015}.

We start all the simulations at redshift $z_{ini}=20$, generating the initial conditions using second-order Lagrangian perturbation theory (2LPT), from an input power spectrum obtained using \texttt{CAMB}\footnote{\href{https://github.com/cmbant/CAMB}{https://github.com/cmbant/CAMB}} \citep{CAMB}. We run pairs of simulations under \LCDM\ and $f(R)$ gravity, each possessing the same cosmological parameters -- except the strength of modified gravity \fRnot.  The initial phases are drawn from a random distribution: we change the initial seed for every $\{$\LCDM$,f(R)\}$ pair of simulations.

\begin{table}
	\centering
	\caption{Calibration of the screening efficiency. The quasi-N-body code \fml{} uses the screening efficiency \citep[see Eq.~(2.6)]{Winther_2015,FML} to characterise the effects of modified gravity efficiently and numerically fast, without solving the full-nonlinear equations behind a theory. In this work, we calibrate the screening efficiency so as to match the boost predicted by \texttt{eMantis} \citep{eMantis}, using the values reported in this Table.}
	\label{tab:se}
	\begin{tabular}{cc}
	$\log_{10} |f_{R_0}|$ & screening efficiency \\
	\hline
	$-$4	& 4 \\
	$-$4.5 	& 2.6 \\
	$-$5	& 2.3 \\
	$-$5.5	& 2.5 \\
	$-$6	& 2.6 \\
	$-$6.5	& 3.5 \\
	$-$7	& 8.2 \\
	$-$7.5	& 8.2 \\
	\end{tabular}
\end{table}

The \fml{} code uses an efficient and fast method to characterise nonlinear modified gravity equations by means of a parameter called \textit{screening efficiency} \citep[see Eq.~(2.6)]{Winther_2015,FML}. This parameter can be calibrated to match the boost of a target simulation or emulator. In this work, we calibrate the screening efficiency against the \texttt{eMantis}\footnote{\href{https://gitlab.obspm.fr/e-mantis/e-mantis}{https://gitlab.obspm.fr/e-mantis/e-mantis}} code \citep{eMantis}, using the values reported in Table~\ref{tab:se}. Note that \texttt{eMantis} is trained on simulations obtained with the AMR code \texttt{ECOSMOG}, which is one of the most accurate in its kind. Further details about how this calibration was chosen are given in Appendix~\ref{appendix:eMantis}. In Figure~\ref{fig:se_boxplot}, we compare the boost obtained from our simulations with that emulated with \texttt{eMantis}: we find $\sim1-2\%$ agreement up to $k\sim1$ $h$/Mpc.

The approximation of using a screening efficiency is fast and accurate over several orders of magnitude in \fRnot, but it is not designed to reproduce arbitrarily fine differences in modified gravity strength: it is for that reason that we restrict ourselves to the discrete \fRnot values mentioned, although we test our emulator on intermediate values in Appendix \ref{appendix:frac_fR0}.

The effects of modified gravity are not equally easy to learn for large or small values of the modified gravity parameter $|f_{R_0}|$: in fact, small values like $\log_{10}|f_{R_0}|=-7.5$ only generate a boost of $\sim10^{-4}$ in the matter power spectrum. For such a model, the penalty for returning the input \LCDM\ snapshot as prediction would be very small.

On the other hand, smaller values of \fRnot are the most important to test for observationally, in case the true law of gravity manifests itself as a small departure from general relativity, which is the most likely scenario given current observational constraints \citep{koyama, Desmond}. This means that models with very small \fRnot are those we may wish to characterise more accurately. Conversely, larger values of \fRnot -- despite being observationally uninteresting, in that they are ruled out -- are very valuable in training the network on the real effects of modified gravity.

For that reason, we build the training set so that larger values of the strength of modified gravity are better represented, as follows. Let us define:
\begin{equation} \label{eq:split_1}
	m(f_{R_0}) = \log{\left|f_{R_0}\right|} -\min{\left(S_{|f_{R_0}|}\right)} + c,
\end{equation}
where $c=0.3$ is an offset we determined empirically, and $\min{\left(S_{|f_{R_0}|}\right)} =-7.5$ is the minimum value we consider for $\log$\fRnot. Then the number of simulations we generate for each \fRnot value is:
\begin{equation} \label{eq:split_2}
	\text{num}(f_{R_0}) = \left\lfloor \frac{m(f_{R_0})}{\sum_{\log{\left|f_{R_0}\right|}} m(f_{R_0})}  \times 1600 \right\rceil
\end{equation}
where $1600$ is the total number of simulations in the training set and $\left\lfloor \cdot \right\rceil$ is the rounding function.

Conversely, we split the validation and test sets equally among the values of \fRnot, to allow us to assess how well the machine-learning model has effectively learnt to perform its task across \fRnot values. The results of this split are reported in Table~\ref{tab:split}.

\begin{table}
	\centering
	\caption{How we divide the $2000$ simulations in the training, validation and test datasets among the eight values of \fRnot we consider, reported in the first column. In the $1600$ simulations comprising the training set, larger values of \fRnot are better represented to help train the network on the effects of modified gravity, following Eqns.~\eqref{eq:split_1}-\eqref{eq:split_2}. Instead, the $400$ simulations making up the combined validation and test sets are equally split among the different strengths of modified gravity. This helps us asses how well the network has learned the mappings in Eqns.~\eqref{eq:displacement}-\eqref{eq:velocity}.}
	\label{tab:split}
	\begin{tabular}{lcccc} 
		\hline
		$\log_{10} |f_{R_0}|$ & \makecell{num sims \\ training} & \makecell{num sims \\ validation} & \makecell{num sims \\ test} & total \\
    		\hline
		$-$4    & 371 & 25 & 25 & 421 \\
		$-$4.5 & 322 & 25 & 25 & 372 \\
		$-$5    & 273 & 25 & 25 & 323 \\
		$-$5.5 & 224 & 25 & 25 & 274 \\
		$-$6    & 176 & 25 & 25 & 226 \\
		$-$6.5 & 127 & 25 & 25 & 177 \\
		$-$7    & 78  & 25 & 25 & 128 \\
		$-$7.5 & 29  & 25 & 25 & 79   \\
		\hline
		total & 1600 & 200 & 200 & 2000 \\
	\end{tabular}
\end{table}

\subsection{Machine learning model}

In this work we use the U-Net/V-Net \citep{Ronneberger2015, Milletari2016} model architecture presented in \cite{map2map}. In the context of this work, the goal of this architecture is to downsample and then upsample simulations using multiple convolutional layers, allowing the network to learn the style parameters that transform \LCDM\ simulations into modified gravity simulations. The characteristic `skip connections' in the U-Net/V-Net architecture between corresponding sampling levels ensure that higher resolution details are preserved during the sampling process. For specific details on the network layers, including convolution parameters and activation functions, we refer interested readers to \cite{map2map}.

Given that we are using the \mtm\ architecture and similar hardware, it follows that we have similar restrictions, such as not being able to process one entire simulation of $512^3$ particles in the V-Net at once. It is for this reason that we also use crops of the simulations (made up of $128^3$ particles) as in \cite{map2map}. Separately, we also follow their padding convention, designed to preserve translational invariance.

Convolutional layers -- such as the ones used in this work -- are excellent at detecting local features, but their predictive power is limited to the extent of their receptive fields -- about $200$ Mpc$/h$ in our case. Considering that the model trains on crops of simulations, we would expect the network to perform better on smaller scales than larger ones. However, for scale-dependent modified gravity models, like the $f(R)$ theory we are considering, the \LCDM\ prediction provides a good description at the large scales, meaning that the network only needs to focus on learning the quasi- and nonlinear scales\footnote{If the scale dependence reached or crossed into the scale of the network receptive field, its architecture would need modifying. We will show in Sec.~\ref{sec:Results} that our recovery of the large scales is excellent, so that we do not need to adapt it.}.

Following \cite{map2map}, we use loss functions aiming to preserve both the Eulerian and Lagrangian properties of a snapshot. In principle, if the network learned the particles' positions or velocities with perfect accuracy, that would guarantee that nonlinear properties like the density or momentum fields also be learned equally well. However, for a finite number of iterations, and in practice, even small errors in the Lagrangian positions or velocities can result in large inaccuracies in the reconstructed Eulerian fields - for instance, in the form of the wrong amount of power at some scales. Incorporating Eulerian elements in the loss greatly alleviates this problem \citep{map2map}.

For the the $\boldsymbol{\Psi}^{(\Lambda \mathrm{CDM})} \mapsto \boldsymbol{\Psi}^{(\mathrm{MG})}$ mapping, we use the loss function:
\begin{equation}
	\log{\mathcal{L}} = \log{\mathcal{L}_{\delta(\mathbf{x})}} + \lambda\log{\mathcal{L}_{\boldsymbol{\Psi}(\mathbf{q})}},
\end{equation}
where $\mathcal{L}_{\delta(\mathbf{x})}$ and $\mathcal{L_{\boldsymbol{\Psi}(\mathbf{x})}}$ are the mean square error (MSE) of the (Eulerian) number count $n(\mathbf{x})$ and the (Lagrangian) displacement $\boldsymbol{\Psi}(\mathbf{q})$, respectively, and $\lambda$ is a parameter regulating the trade-off between the Eulerian and Lagrangian losses. Empirically, we find that $\lambda=3$ gives the best results: we can explain this intuitively by noting that the network needs to learn three real numbers to reproduce $\boldsymbol{\Psi}$, versus a single real number for $n$; because the MSE compresses snapshot-wide losses to a single real number, the simple sum of log-MSE losses results in relative downweighting of the original Lagrangian loss.

For the velocity network,  we employ a loss function that combines isotropic and anisotropic properties of the Eulerian velocity field, as well as the Lagrangian loss like before. We find that reproducing anisotropic features of the Eulerian velocity field is essential in order to predict the redshift-space distortions (RSD) accurately. Having defined $\mathcal{L}_{\mathbf{v}(\mathbf{q})}$ and $\mathcal{L}_{\mathbf{v}(\mathbf{x})}$ as the loss on the Lagrangian and Eulerian velocity fields similarly to before, and having further defined $\mathcal{L}_{\mathbf{v}(\mathbf{x})^2}$ as the MSE loss on the quantity:
\begin{equation}
	\mathcal{L}_{\mathbf{v}(\mathbf{x})^2} = m  \sum_{ij} v(\mathbf{x})^i v(\mathbf{x})^j ,
\end{equation}
the overall loss we employ for the $\mathbf{v}^{(\Lambda \mathrm{CDM})} \mapsto \mathbf{v}^{(\mathrm{MG})}$ training is:
\begin{equation}
	\log \mathcal{L}  = \log \mathcal{L}_{\mathbf{v}(\mathbf{q})}  + \log \mathcal{L}_{\mathbf{v}}(\mathbf{x}) + \log \mathcal{L}_{\mathbf{v}(\mathbf{x})^2} .
\end{equation}
We note that in order to compute $\mathbf{v}(\mathbf{x})$, we need a displacement field to obtain $\mathbf{x}(\mathbf{q};t) = \mathbf{q} + \boldsymbol{\Psi}(\mathbf{q}; t)$. Therefore, the network for the velocity mapping, Eq.~(\ref{eq:velocity}), requires an input displacement. Unlike in the original work by \cite{map2map}, which used the input \LCDM\ displacements for this purpose, we use the outputs from the displacement mapping in Eq.~(\ref{eq:displacement}); in other words, we first train the displacement network and then use its output to compute the loss function in the training of the velocity mapping. Although our choice does not allow the two networks to be trained independently of each other, we find that it results in significant improvement in the accuracy of velocity-based results, because of the effectiveness with which our displacement network is able to reproduce the target displacement (see Sec.~\ref{sec:Results}). 

We express input and target displacements in units of Mpc$/h$ and input and target velocities in $\mathrm{km}/\mathrm{s}$. The style parameters are normalised as:
\begin{equation}
	s(f_{R_0}) = \frac{ \log_{10}\left| f_{R_0} \right| - \mu\left( S_{|f_{R_0}|} \right) } { \Delta( S_{|f_{R_0}|} ) },
\end{equation}
where $\mu\left( S_{|f_{R_0}|} \right)$ is the mean of the \fRnot values, and $\Delta( S_{|f_{R_0}|} )$ is the maximum deviation from that mean.

\subsection{High Performance Computing implementation}
We built upon the existing \mtm\ codebase to ensure it worked with our \texttt{Sciama} HPC cluster \footnote{\href{https://sciama.icg.port.ac.uk/}{https://sciama.icg.port.ac.uk/}} and was compatible with the latest python packages, as well as making several modifications and improvements.

Due to the large amount of data that needs to be communicated across compute nodes during training, considerations must be made regarding \mtm's data pipeline efficiency.
Extensive benchmarking of the original map2map codebase was carried on our \texttt{Sciama} HPC cluster to estimate the overall training time and identify bottlenecks. By using the comprehensive set of inbuilt benchmarking tools in \texttt{PyTorch}, we were able to identify some limitations caused by our \texttt{Sciama} HPC cluster configuration. This meant that we were not able to use NVIDIA's Direct Memory Access (NVIDIA GPUDirect) technology\footnote{\href{https://developer.nvidia.com/gpudirect}{https://developer.nvidia.com/gpudirect}}, which allows for faster data transfer between GPUs.
To account for this limitation, extra optimisations were made to the \mtm\ codebase to reduce the overall training time to an acceptable level. 

We modified \mtm\ to include Automatic Mixed Precision\footnote{\href{https://pytorch.org/docs/stable/amp.html}{https://pytorch.org/docs/stable/amp.html}} (AMP) training, which allows for faster training times by using a lower precision datatype like \texttt{float16} for the majority of the training process, and only using a higher precision one like \texttt{float32} when necessary. Through empirical testing, we have set AMP to use the \texttt{bfloat16} datatype, which allows for more dynamic range, but less precision compared to the \texttt{float16} datatype.
This leads to a reduction in the memory footprint and a speed-up in the training process, but also allows for parts of the training to leverage the Tensor Cores on the NVIDIA A100 GPUs, which can perform matrix multiplications at a much faster rate than the standard CUDA cores. \texttt{PyTorch}'s Just-In-Time (JIT) compiler was used to compile the model code before training, which allows for faster model inference. We also refactored the code to use \texttt{PyTorch}'s new \texttt{torchrun} convention, which improves fault-tolerance and allows for better scaling across multiple GPUs\footnote{\href{https://pytorch.org/tutorials/beginner/ddp\_series\_fault\_tolerance.html}{https://pytorch.org/tutorials/beginner/ddp\_series\_fault\_tolerance.html}}.

Other optimisations and improvements were also investigated and tested, such as using a batch size greater than $1$, gradient checkpointing, gradient accumulation, fusing the optimizer and backward pass into one calculation, and Fully Sharded Data Parallel (FSDP) mode, all of which are detailed on the \texttt{Pytorch} website. However, these were not implemented in the final version of the code after it was decided that the inclusion of AMP and JIT compilation was sufficient to achieve the desired training times. These additional optimisation methods should be considered for future \mtm{} calculations.

The cluster resources used in this work consist of 3x GPU compute nodes, each with 2x AMD EPYC 7713 processors (128 CPU cores), at least 512Gb RAM, and two NVIDIA A100 GPUs (with at least 40Gb VRAM), with the simulations stored on a Lustre filesystem. Each GPU was able to process one simulation at a time, giving an effective batch size of $6$ when distributed across the $6$ total NVIDIA A100 GPUs. The Adam optimiser \citep{Adam} was used with a learning rate of $10^{-5}$ and hyperparameters $\beta_1=0.9$, $\beta_2=0.999$, and the learning rate reduces by 10x after the training loss stops decreasing for two iterations. We monitored the validation loss, and examined the accuracy of summary statistics extracted from the validation dataset, to determine when to stop training the network; as a result we trained the mapping in Eq.~\ref{eq:displacement} for $76$ iterations, and the one in Eq.~\ref{eq:velocity} for 109 iterations. The training process took approximately 444 hours to complete for the displacement network and 917 hours for the velocity network.

\section{Results} \label{sec:Results}
 
In this section, we present the performance of our emulator on the test set, i.e. a dataset entirely comprised of simulations that the algorithm did not see during training, and that were not used when validating results.
 
In Figure~\ref{fig:m2m_vs_real}, we compare the density field reconstructed from one emulated snapshot against its target counterpart -- in this example, \fRnot$=10^{-5}$. In particular, we show one same slice $250$ Mpc$/h$ $\times$ $250$ Mpc$/h$ wide and $29.3$ Mpc$/h$ deep, extracted from both simulations. The left and middle columns show the emulated and target overdensity, expressed as $1 + \delta$, whereas the rightmost column shows the relative error on the emulator's output. We can see that the emulator is capable of reproducing the cosmic web faithfully, and characterising correctly not only the amplitude, but also the phases of the density field.  
 
This section is organised as follows. In Sec.~\ref{sec:Summary_statistics}, we extract several summary statistics from the output and target $f(R)$ snapshots, and compare them in Figures~\ref{fig:results_density_displacement}-\ref{fig:results_stochasticities}. We continue by comparing the modified gravity boost -- i.e. $P(k)_{f(R)}/P(k)_{\Lambda\mathrm{CDM}}$, where $P(k)$ is the matter power spectrum -- obtained from the emulated and \LCDM\ snapshots against the prediction from \texttt{eMantis}, showing our results in Figure~\ref{fig:comparison_w_eMantis}.
In Sec.~\ref{sec:comparison_against_rescaling}, we compare the output of our emulator against the algorithm proposed by \cite{Mead_2015}, which uses rescaling techniques. 
Finally, in Sec.~\ref{Sec:extrapolations}, we evaluate the network's performance in extrapolation, by testing the emulator on cosmological parameters it was not trained on. In particular, we focus on varying $\Omega_M, \sigma_8$ and $h$, as well as considering the parameters of the Euclid reference simulations \citep{EE2}. We present the result in Figure~\ref{fig:results_extrapolations}.

\subsection{Performance on summary statistics} \label{sec:Summary_statistics}

In this Section, we show the performance of the emulator as measured by several summary statistics that are widely used. 

Because our training, validation and test datasets are made up of dark-matter-only simulations, we display any quantity computed from the density and displacement fields only up to scales \(k \sim 1 \, h/\mathrm{Mpc}\), to avoid contamination from baryonic effects.

Separately, and to avoid being affected by uncertainties in the galaxy-halo connection, we quote any results calculated from the Eulerian momentum or Lagrangian velocity fields (in particular, the multipoles of the redshift-space distortions) only up to scales $k \sim 0.3 \, h/\mathrm{Mpc}$. Note that the redshift-space distortions smear small-scale nonlinearities to larger scales along the line of sight, due to the ``fingers of God'' effects \citep{Jackson_1972}. This justifies comparing the redshift-space density power multipoles on much larger scales than the real-space density power spectrum.

Except when computing the power spectra and cross-correlations of vector quantities (i.e. Eqns.~\eqref{eq:vector_power_spectrum}-\eqref{eq:cc_coeff_vec}, for which we use a custom code), we compute all summary statistics showed in this section using the \texttt{Pylians}\footnote{\href{https://pylians3.readthedocs.io}{https://pylians3.readthedocs.io}} library \citep{Pylians}.

In all Figures~\ref{fig:results_density_displacement}-\ref{fig:comparison_w_eMantis},  lines are coloured with shades of red proportional to $\log{\left| f_{R_0} \right|}$ (as indicated in the colour bar), highlighting the dependence of the summary statistics (and their error) on the strength of the modified gravity parameter/style parameter. At every scale $k$ and within one same \fRnot value, we further split the predicted power spectra and errors into 100 percentiles, and assign line opacity accordingly: the greatest opacity is given to the 50th percentile (i.e. the median) and progressively decreases towards the 0th/100th percentile.

In all the figures shown in this Section, we focus on values of \fRnot logarithmically spaced by $0.5$ (see Sec.~\ref{sec:datasets}); a comparison on intermediate values is shown in Appendix~\ref{appendix:frac_fR0}, where we show the same level of agreement.

\subsubsection{Power spectra}
 
 \begin{figure*}
 	\centering
	\includegraphics[width=\textwidth]{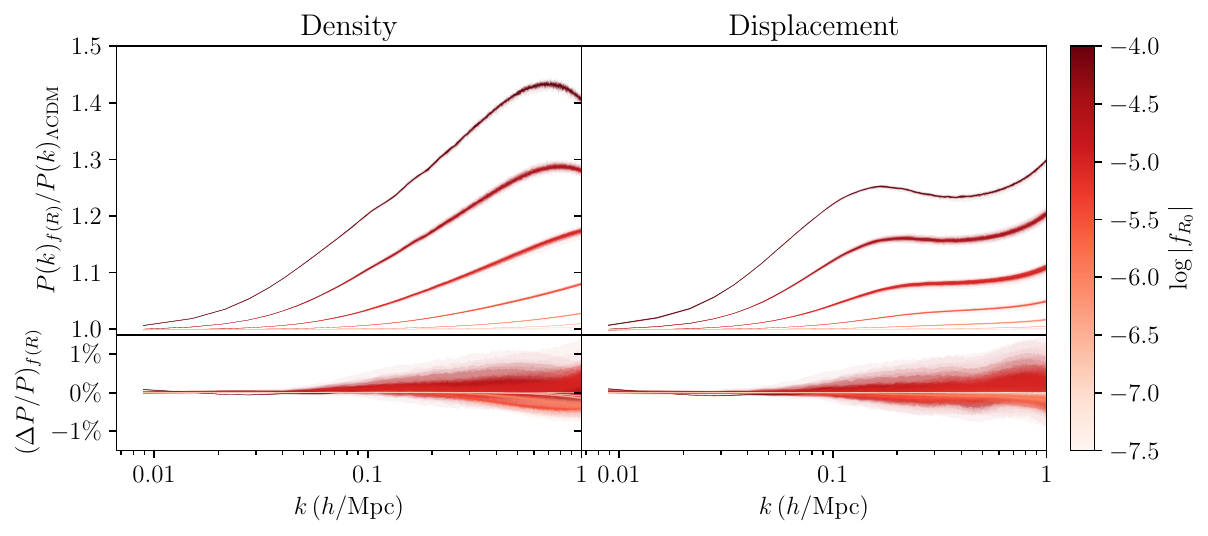}
	\caption{\textit{Top panels:} the boost factor \textit{(left)} and the ratio of the power spectra of the displacement field in $f(R)$ and \LCDM\ \textit{(right)}. The power spectrum of the displacement field is defined in Eq.~\eqref{eq:vector_power_spectrum}. \textit{Bottom panels:} the fractional error on the density and displacement power spectra as computed from the snapshot produced by our emulator. In all plots, lines are coloured with shades of red proportional to $\log{\left| f_{R_0} \right|}$, as indicated in the colourbar. At every scale $k$ and within one same \fRnot value, we split the predicted power spectra and errors into percentiles, and assign line opacity accordingly: the greatest opacity is given to the 50th (i.e., median) percentile and progressively decreases towards the 0th/100th percentile. We find an accuracy of $\sim 1\%$ or better for both power spectra at all scales $k < 1 \, h/\mathrm{Mpc}$, in line with the requirements of Stage IV experiments. 
 }
	\label{fig:results_density_displacement} 
 \end{figure*}
 
 \begin{figure*}
 	\centering
	\includegraphics[width=\textwidth]{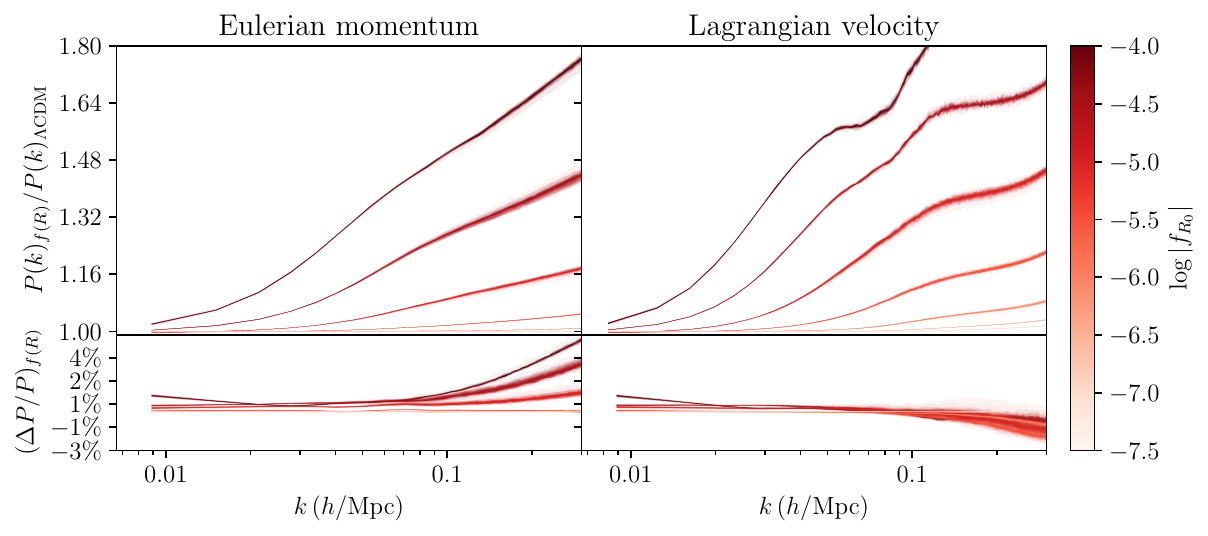}
	\caption{\textit{Top panels:} Ratio of the power spectra of the Eulerian momentum \textit{(left)} and Lagrangian velocity fields \textit{(right)}, in $f(R)$ and \LCDM{}. The power spectrum for these fields is defined in Eq.~\eqref{eq:vector_power_spectrum}. \textit{Bottom panels:} the fractional error on the same power spectra, when computed from the snapshot produced by our emulator. 
 We find an agreement of $\sim 4\%$ or better at all scales $k < 0.3 \, h/\mathrm{Mpc}$, which are the scales of relevance in the analysis of the redshift-space distortions.}
	\label{fig:results_eul_v_lag_v}
 \end{figure*}

First we consider the density power spectrum, defined as
    \begin{equation}
    	\langle \delta(\mathbf{k}_1) \delta(\mathbf{k}_2) \rangle = {(2\pi)}^3 \delta_D ( \mathbf{k}_1 + \mathbf{k}_2 ) P(k),
    \end{equation}
where angular brackets denote ensemble average, $\delta_D$ is the Dirac delta function, and $\mathbf{k}_1,\mathbf{k}_2$ are Fourier modes.
    
For vector fields like the displacement, Eulerian momentum, and Lagrangian velocity fields, we can sum the power spectrum over spatial dimensions, e.g.:
     \begin{equation} \label{eq:vector_power_spectrum}
    	\langle \boldsymbol{\Psi}(\mathbf{k}_1) \cdot \boldsymbol{\Psi} (\mathbf{k}_2) \rangle = {(2\pi)}^3 \delta_D( \mathbf{k}_1 + \mathbf{k}_2) P_{\boldsymbol{\Psi} \boldsymbol{\Psi}}(k),
    \end{equation}
 where the symbol $\cdot$ indicates the Euclidean dot product.
 
 As indicated in Sec.~\ref{sec:mapping}, we reconstruct the Eulerian density and momentum from the particles' displacements and velocities using the CIC algorithm.
 
 In Figures~\ref{fig:results_density_displacement}-\ref{fig:results_eul_v_lag_v}, we show the performance of our emulator as measured by the power spectrum of the density, displacement, Eulerian momentum and Lagrangian velocity fields. In the top panels, we present the target statistics as ratios to the corresponding \LCDM\ quantities: we compute them from the target and input simulations, respectively. For the density power spectrum, this is the widely-studied modified-gravity (MG) boost $P(k)_{\mathrm{MG}} / P(k)_{\Lambda\mathrm{CDM}}$. In the bottom panels, we display the fractional error on the four summary statistics, as computed from the emulated snapshots and compared against the target.
 
 For the density and displacement power spectra, we find an accuracy of $\sim 1\%$ or better at all scales \(k < 1 \, h/\mathrm{Mpc}\), in line with the requirements of Stage IV experiments. 
 For the Eulerian momentum and Lagrangian velocity power spectra, we find an accuracy of $\sim 4\%$ or better at scales  \(k < 0.3 \, h/\mathrm{Mpc}\), which are the scales of relevance in the analysis of the redshift-space distortions.

 \subsubsection{Bispectrum of the density field}
\begin{figure*}
 	\centering
	\includegraphics[width=\textwidth]{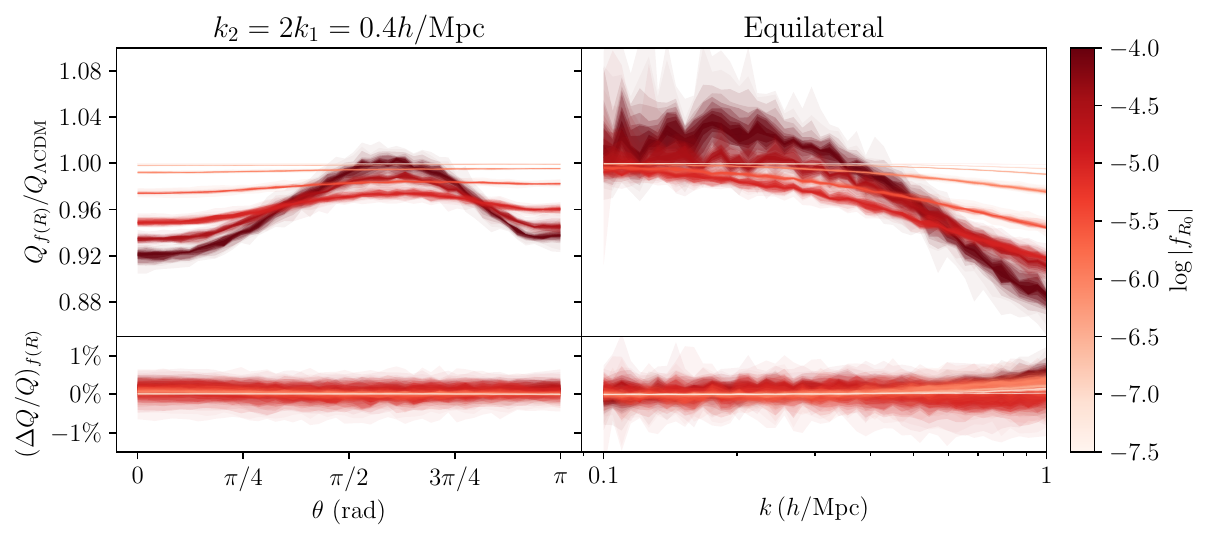}
	\caption{\textit{Top panels:} $f(R)/$\LCDM\ ratio of the reduced bispectra in the configuration $k_2=2k_1=0.4 h$/Mpc \textit{(left)} and in the equilateral configuration \textit{(right)}. \textit{Bottom panels:} the fractional error on the same bispectra,as computed from the output snapshots produced by our emulator. 
 We find an agreement of $\sim 0.8\%$ or better at all angles in the configuration $k_2=2k_1=0.4 h$/Mpc, and $\sim 1.5\%$ or better at all scales $k < 1 h/$Mpc in the equilateral configuration.}
	\label{fig:results_bispectrum}
 \end{figure*}
 
By emulating the full snapshot, we can obtain statistics beyond the power spectrum that are able to capture any non-Gaussianities in the density field. The leading order statistics for this purpose is the bispectrum, which is defined as:
    \begin{equation}
    	\langle \delta(\mathbf{k}_1) \delta(\mathbf{k}_2) \delta(\mathbf{k}_3) \rangle = {(2\pi)}^3 \delta_D( \mathbf{k}_1 + \mathbf{k}_2+ \mathbf{k}_3) B(k_1, k_2).
    \end{equation}
Unlike the power spectrum, which is only sensitive to the magnitude of Fourier modes, the bispectrum is the lowest-order correlator that is sensitive to phases.

Because homogeneity constraints the wavenumbers $(\mathbf{k}_1 + \mathbf{k}_2 + \mathbf{k}_3)$ to form a closed triangle, we can also express the bispectrum as a function of two magnitudes and an angle, i.e. $B(k_1,k_2,\theta)$, which we do in Figure~\ref{fig:results_bispectrum}. 
It is useful, particularly in analyses of modified theories of gravity, to consider the \textit{reduced bispectrum}:
   \begin{equation} 
   	Q(k_1, k_2, k_3) = \frac{B(k_1, k_2, k_3)}{P(k_1)P(k_2) + P(k_2)P(k_3)+P(k_1)P(k_3)} ,
   \end{equation}
to remove the information that is already contained in the power spectrum. 

In Figure~\ref{fig:results_bispectrum}, we show $Q(k_1,k_2,\theta)$ for the equilateral configuration (right panels) and for $k_2=2k_1=0.4 h$/Mpc (left panels), to allow us to compare our results with \cite{Gil_Mar_n_2011}. 
As before, we display the $f(R)/$\LCDM\ ratio of the examined quantities in the top panels, and the relative error in the prediction computed from the emulated snapshots in the bottom panels. We find an agreement of $\sim 0.8\%$ or better at all angles in the configuration $k_2=2k_1=0.4 h$/Mpc, and $\sim 1.5\%$ or better at all scales $k < 1 h/$Mpc for the equilateral configuration. 

We note that the screening approximation used in \fml{} simulations has some impact on the bispectrum: in \fml{}, the scalar field equation is linearised for speed, which results in the bispectrum missing some contributions from the non-linearity of the scalar field \citep{Fiorini2023}. For \fRnot $=10^{-5}$, the effect is largest, and \fml{} simulations produce a bispectrum signal consistent with the results of \cite{Gil_Mar_n_2011} without the chameleon mechanism. The effect of screening is instead smaller for \fRnot $=10^{-4}$ and \fRnot $=10^{-6}$ \citep{Gil_Mar_n_2011}, where we achieve greater accuracy. However, when investigating the impact of the \fml{} screening approximation on the galaxy clustering bispectrum, which is actually observable, \cite{Fiorini2023} also showed that the bispectrum is dominated by the nonlinear galaxy bias, so that inaccuracies in the screening approximation become comparatively negligible.

\subsubsection{Redshift-space distortions (RSD)}

 \begin{figure*}
 	\centering
	\includegraphics[width=\textwidth]{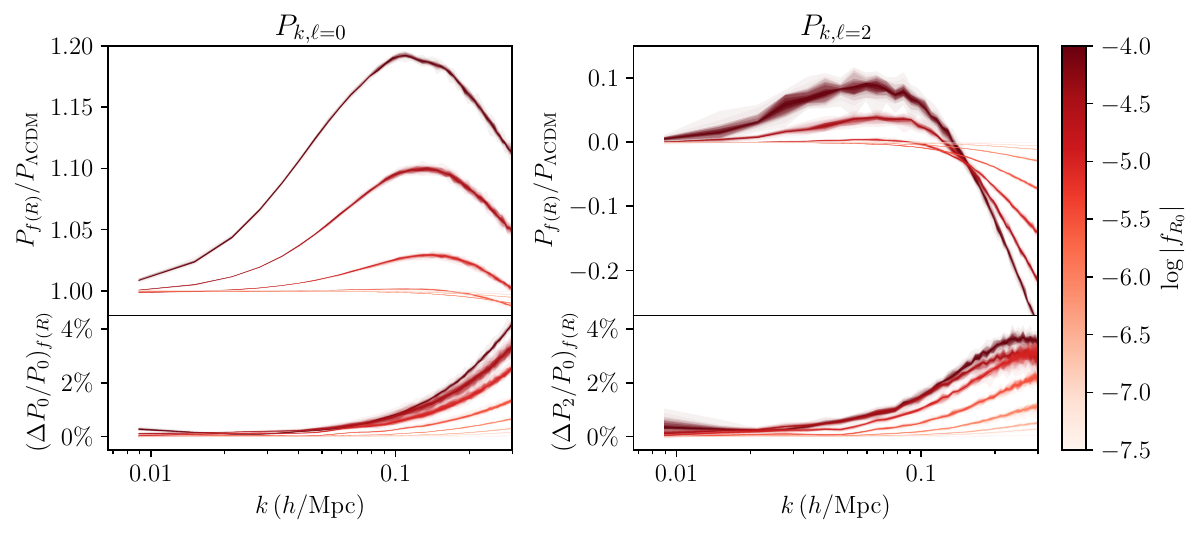}
	\caption{\textit{Top panels:} the ratio of the monopole \textit{(left)} and quadrupole \textit{(right)} of the redshift-space distortions between $f(R)$ and \LCDM. \textit{Bottom panels:} fractional error on the monopole \textit{(left)} and quadrupole \text{(right)} as computed from the emulated snapshot (positions and velocities). Because the quadrupole crosses zero at $k \sim 0.1-0.2$ $h$/Mpc, we normalise the error on the quadrupole by the magnitude of the monopole. 
 We find $\sim 4\%$ agreement or better in both the monopole and the quadrupole at all scales $k < 0.3 h/$Mpc.}
	\label{fig:results_RSD}
 \end{figure*}
 
The peculiar velocities of galaxies introduce anisotropies in the observed galaxy distribution, in a way that is sensitive to the response of nonrelativistic matter to gravity. Because only radial velocities are observed, the line of sight is selected as a special direction, along which we can decompose the redshift-space power spectrum in Legendre polynomials $\mathscr{L}$:
   \begin{equation}
   	P(k) \sum_{\ell} (2\ell +1 ) P_{\ell}(k) \mathscr{L}(\hat{\mathbf{k}} \cdot \hat{\mathbf{n}})
   \end{equation}
where $\hat{\mathbf{n}}$ is the line of sight and $\,\hat{}\,$ indicates taking a unit vector.

In Figure~\ref{fig:results_RSD}, we show the performance of our emulator in predicting the monopole ($\ell=0$) and quadrupole ($\ell=2$) of the redshift-space distortions. As in previous figures, we show the magnitude of the effects we are trying to capture, visualised through the $f(R)$/\LCDM{} ratio, in the top panels; in the bottom panels we show the fractional error on the same quantities as computed from the emulated snapshots. However, unlike in previous figures, we normalise the absolute error on the quadrupole by the monopole, i.e.: $(P_{\ell=2, \mathrm{out}} - P_{\ell=2, \mathrm{true}}) / P_{\ell=0, \mathrm{true}}$. This is because the quadrupole becomes zero around $k \sim 0.1-0.2$ $h$/Mpc, making the relative error $(P_{\ell=2, \mathrm{out}} - P_{\ell=2, \mathrm{true}}) / P_{\ell=2, \mathrm{true}}$ difficult to interpret.

We find $\sim 4\%$ agreement or better in both the monopole and the quadrupole at all scales $k < 0.3 h/$Mpc.

\subsubsection{Stochasticities} 

 \begin{figure*}
 	\centering
	\includegraphics[width=\textwidth]{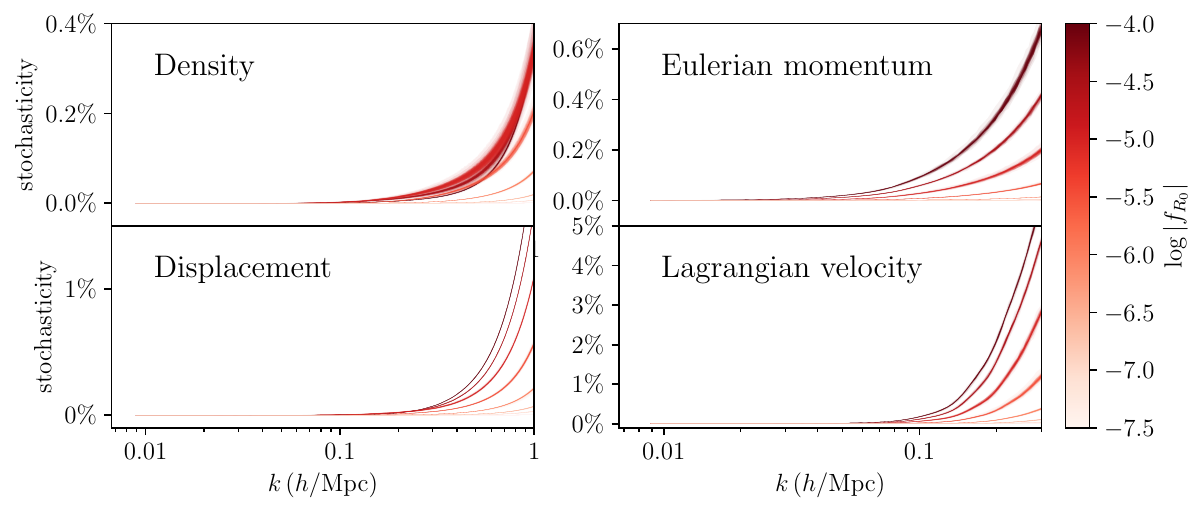}
	\caption{Stochasticity of the density \textit{(top left)}, displacement \textit{(bottom left)}, Eulerian momentum \textit{(top right)} and Lagrangian velocity \textit{(bottom right)}. The stochasticity is defined as $1-r^2$, where $r$ is the cross-correlation coefficient defined in Eqns.~\eqref{cc_coeff}-\eqref{eq:cc_coeff_vec}. The stochasticity is sensitive to phases and quantifies the variance in the emulated snapshot that is unaccounted for in the target snapshot. 
 For the density and displacement fields, we find stochasticities less than $\sim 0.4\%$ and $\sim 2\%$, respectively, over the scale range $k < 1$ $h/$Mpc. For the Eulerian momentum and the Lagrangian velocity fields, we find stochasticities less than $0.6\%$ and less than $5\%$, respectively, over the scale range $k < 0.3$ $h/$Mpc. Overall, these results indicate excellent agreement.}
	\label{fig:results_stochasticities}
 \end{figure*}

Let us define the cross-correlation coefficient of the density field as
    \begin{equation} \label{cc_coeff}
    	r(k) = \frac{ \langle \delta_{\mathrm{out}}(\mathbf{k}) \delta_{\mathrm{true}}(\mathbf{k}^{\prime})\rangle }{ \sqrt{\langle \delta_{\mathrm{out}}(\mathbf{k}) \delta_{\mathrm{true}} (\mathbf{k}^{\prime})\rangle} \sqrt{\langle \delta_{\mathrm{true}}(\mathbf{k}) \delta_{\mathrm{true}} (\mathbf{k}^{\prime}) \rangle} }.
    \end{equation}
Similarly, we define it for the Eulerian momentum (which is a vector field) as:
       \begin{equation} \label{eq:cc_coeff_vec}
    	r(k) = \frac{ \langle \mathbf{P}_{\mathrm{out}}(\mathbf{k}) \cdot \mathbf{P}_{\mathrm{true}}(\mathbf{k}^{\prime})\rangle }{ \sqrt{ \langle \mathbf{P}_{\mathrm{out}}(\mathbf{k}) \cdot \mathbf{P}_{\mathrm{true}} (\mathbf{k}^{\prime})\rangle} \sqrt{ \langle \mathbf{P}_{\mathrm{true}}(\mathbf{k}) \cdot \mathbf{P}_{\mathrm{true}} (\mathbf{k}^{\prime}) \rangle }},
    \end{equation}
and analogously for the displacement and Lagrangian velocity fields. In these two equations, the pedices ``$\mathrm{out}$'' and ``$\mathrm{true}$'' stand for the emulated and target snapshot.
    
The \textit{stochasticity} $1-r^2$ then quantifies the excess variance in the output snapshot that is unaccounted for in the target snapshot.

In Figure~\ref{fig:results_stochasticities}, we show the stochasticities of the density, displacement, Eulerian momentum and Lagrangian velocity fields: for the density and momentum fields, the stochasticity is less than $\sim 0.4\%$ and $\sim 0.6\%$, respectively, at scales $k < 1$ $h/$Mpc and $k < 0.3$ $h/$Mpc. This indicates very little stochasticity. For the displacement and velocity fields, we still find a stochasticity that is less than $\sim 2\%$ and less than $\sim 5\%$, respectively, over the same scale ranges.

Overall, these results indicate excellent agreements between the emulated and target snapshots: note that the cross-correlation coefficient in Eqns.~\eqref{cc_coeff}-\eqref{eq:cc_coeff_vec} is sensitive to phases.
   
 \subsubsection{Comparison against eMantis} \label{sec:comparison_w_eMantis}
  
\begin{figure}
	\centering
	\includegraphics[width=\columnwidth]{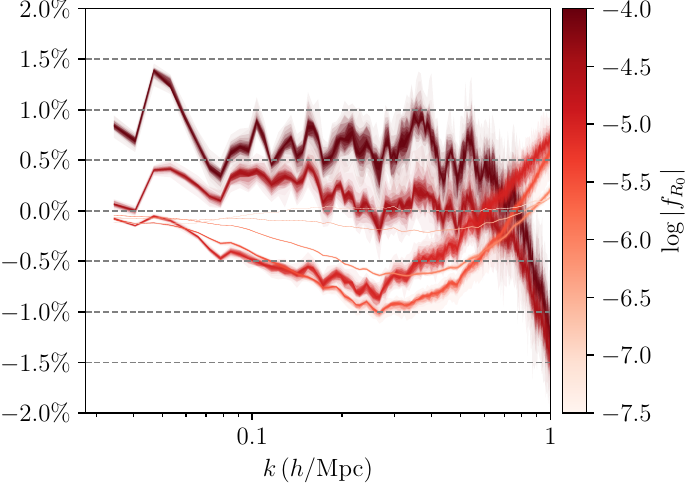}
	\caption{Comparison of the boost factor obtained from the emulated and input \LCDM\ snapshots against the predictions of \texttt{eMantis}. We find an agreement $\sim 1.5\%$ at all scales $k < 1$ $h/$Mpc. 
 }
	\label{fig:comparison_w_eMantis}
\end{figure}

We can compare the boost calculated from the output (and \LCDM{}) snapshots against the prediction from \texttt{eMantis}, giving us the opportunity to test our results against an independent code. This comparison is shown in Figure~\ref{fig:comparison_w_eMantis}.

We obtain the same level of agreement as in calibration -- 
confirming that the network has learned the properties of the target snapshots -- and resulting in an overall agreement of $\sim 1.5\%$ at all scales $k < 1$ $h/$Mpc.

\subsection{Comparison against rescaling methods} \label{sec:comparison_against_rescaling}

\begin{figure*}
\centering
\includegraphics[width=\textwidth]{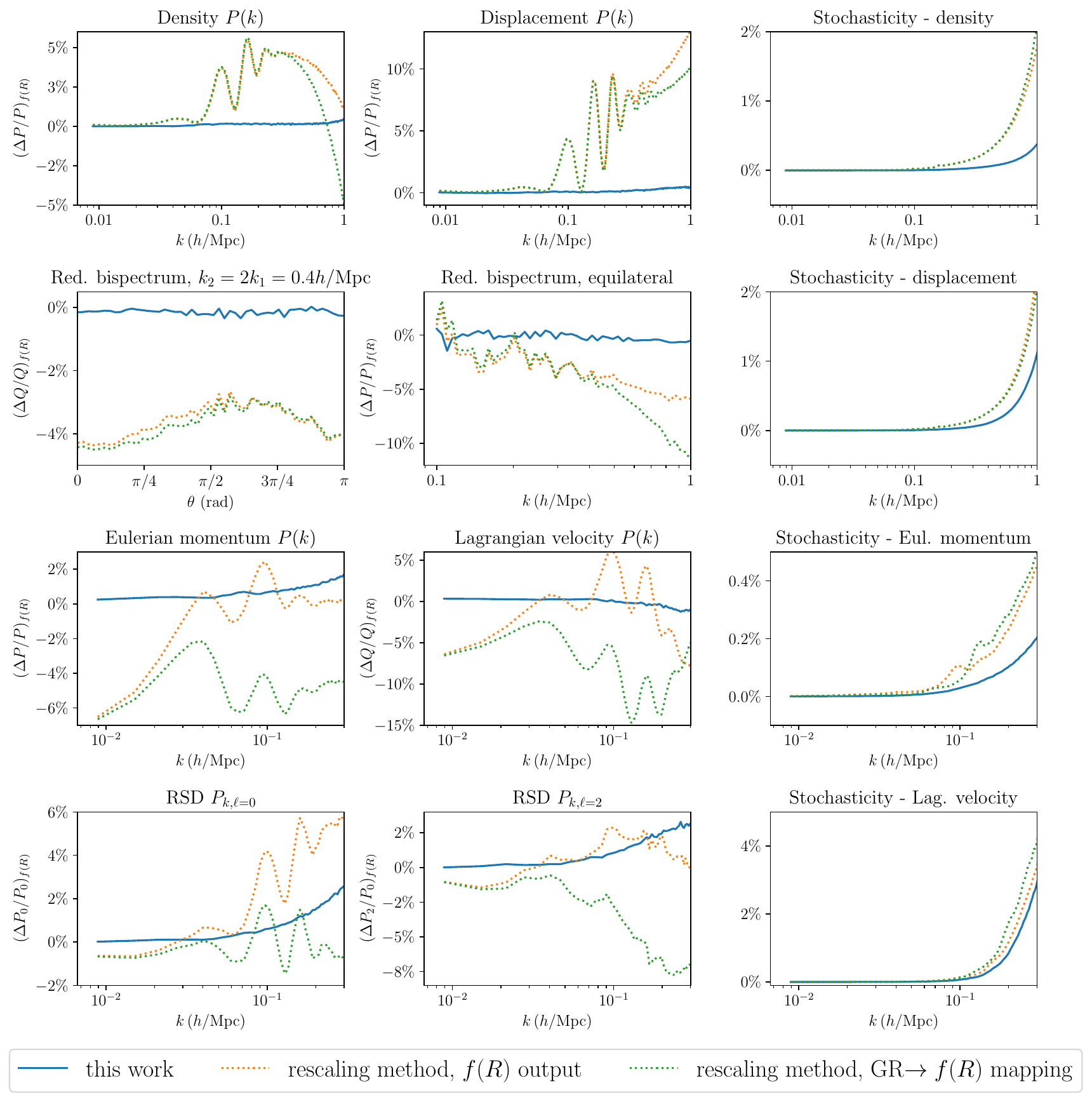}
\caption{The performance of our emulator (solid blue line) against the rescaling method of \citet{Mead_2015}, evaluated on the generation of a $f(R)$ output with \fRnot$=10^{-5}$ (orange dotted line), and in reproducing the mapping from standard to $f(R)$ gravity (dotted green line). Details of how we computed each prediction are given in Sec.~\ref{sec:comparison_against_rescaling}. Our emulator performs considerably better than the earlier method across several summary statistics; in particular, we are not affected by unwanted discrepancies in the baryon acoustic oscillations.}
\label{Fig:comparison_with_rescaling_method}
\end{figure*}

In this Section, we compare the performance of our emulator against the rescaling method introduced by \citet{Mead_2015}, which builds on the approach initially proposed by \citet{Angulo_2010}. This method reproduces the snapshot of a target cosmology (run under standard or modified gravity) by modifying that of a \textit{parent} \LCDM{} simulation with a high clustering amplitude $\sigma_8$ (e.g., $\sigma_8=1.2$ in \citet{Mead_2015}). If $L^{\prime}$ and $z^{\prime}$ are the box size and redshift of the target snapshot, then the parent simulation is chosen to have box size $L = L^{\prime}/s$, and output redshift $z_{\star}$, where $s$ and $z_{\star}$ are parameters obtained by minimising the difference in linear clustering over a specified range of scales. While this transformation successfully reproduces many features of the target simulation, it introduces discrepancies on linear scales, due to mismatches in the baryon acoustic oscillations (BAOs) between the parent and target cosmologies. This effect is particularly noticeable in $f(R)$ gravity, where the scale-dependent nature of growth necessitates a corresponding scale-dependent rescaling to fully address the issue. To mitigate it, a second step is applied in which the Zel{'}dovich approximation is used to update the particle positions and velocities at the linear scales, aligning them with the target cosmology, and allowing for scale-dependent corrections in the $f(R)$ scenario. Finally, haloes are restructured to match chosen concentration-mass relations and velocity dispersions: this third step significantly improves the accuracy of the predicted redshift-space distortions. For a detailed description of this method, we direct the reader to \citet{Mead_2015}, as well as the original \citet{Angulo_2010}.

The comparison between our emulator and the rescaling algorithm can be done in two ways: we can assess how accurately the methods reproduce the target $f(R)$ simulation, or we can consider how well each reproduces the mapping from standard to modified gravity. In this section, we follow both approaches in turn.

To compare the accuracy in reproducing a modified-gravity snapshot, we first apply the algorithm to a target $f(R)$ simulation with \fRnot$=10^{-5}$. We set the clustering amplitude in the parent simulation to $\sigma_8=1.2$ (as in \citet{Mead_2015}), and match the linear clustering over the scales of $3.3$ and $22.7$ $h/$Mpc -- chosen to represent the scales of the smallest and largest haloes we find in the target snapshot\footnote{The halo catalogue was computed with the friends-of-friends algorithm included in the \texttt{FML} library.}. By doing this, we obtain rescaling parameters $s=0.803$ and $z_{\star}= 0.264$. We update the particle positions and velocity in a scale-dependent way, up to the nonlinearity scale of $k_{NL}=0.1452$ $h/$Mpc, following the Zel{'}dovich approximation. We note that the \texttt{COLA} algorithm is not designed to reproduce halo structure well; nevertheless, we apply the same restructuring prescriptions as described in \citet{Mead_2015}.

We compare the output from this work against that from the rescaling algorithm in Figure~\ref{Fig:comparison_with_rescaling_method} (solid blue and dotted orange lines), for several summary statistics. We can see that our emulator outperforms the earlier algorithm in all measures, with the exception of the small scales in the RSD quadrupole. In particular, the density power spectrum is reproduced up to $\sim5\%$ in the rescaling method, whereas we obtain better than $1\%$ at all scales $k < 1$ $h/$Mpc. Likewise, the bispectrum and the stochasticities show an improvement in phases over earlier work: for example, we recover the equilateral reduced bispectrum to $\sim 1\%$, as opposed to $\sim5\%$ in the rescaling method, at all scales $k < 1$ $h/$Mpc.

Next we consider the mapping from general relativity to modified gravity, which is what the emulator presented in this work seeks to learn. For this comparison, we need to predict the mapping under the rescaling method, which in turn requires that we compute a \LCDM\ prediction as input to the mapping (note that the parent simulation is not suitable for this purpose as it has large clustering amplitude by design). Using an input \LCDM\ snapshot as target, we find $s\sim1.000$ and $z_{\star}=0.718$, choosing again $\sigma_8=1.2$ for the parent simulation, and matching linear clustering at the scales of the smallest and largest halo in the target snapshot ($3.3$ and $23.7$ $h/$Mpc). For the nonlinearity scale, we find $k_{NL}=0.1626$ $h/$Mpc. From the $f(R)$ snapshot predicted at the previous step, and the \LCDM\ one just obtained, we compute the displacement $\Delta\boldsymbol{\Psi} = \boldsymbol{\Psi}_{f(R)} - \boldsymbol{\Psi}_{\rm \Lambda CDM}$ and velocity difference $\Delta\mathbf{v} = \mathbf{v}_{f(R)} - \mathbf{v}_{\rm \Lambda CDM}$, which encode the desired GR$\rightarrow f(R)$ mapping as predicted by the rescaling method.

To compare the performance of our emulator and the rescaling method, we add $\Delta\boldsymbol{\Psi}$ and $\Delta\mathbf{v}$ to the positions and velocities of a ``ground truth'' \LCDM\ snapshot computed with \texttt{FML}, obtaining a new predicted $f(R)$ snapshot. This choice has the disadvantage that the answer from the rescaling method will be affected by errors in both the \LCDM\ and $f(R)$ snapshots; on the other hand, this comparison allows us to test our emulator on the mapping we tasked it to learn.

We show the results of this comparison in Figure~\ref{Fig:comparison_with_rescaling_method} (solid blue and dotted green lines). Our algorithm outperforms the earlier work again, with the exception of the small scales in the RSD monopole. The mapping computed with the rescaling method matches the density power spectrum within $5\%$ at scales $k < 1$ $h/$Mpc, against $1\%$ in our emulator. Similarly, the obtained equilateral reduced bispectrum is only accurate to $11\%$ over the same scales, whereas we obtain $1\%$ or better.

Note that, for both methods of comparison, the output of the rescaling method in $f(R)$ shows residual unwanted imprints of the parent BAOs, consistently with Figure~8 of \citet{Mead_2015}. Although this method achieves impressive results considering its simplicity, and has greater interpretability, the comparison presented in this Section shows that our approach significantly improves the accuracy of the output snapshot, making it an improved tool for field-level inference.

\subsection{Performance on extrapolation} \label{Sec:extrapolations}

\begin{figure*}
	\centering
	\includegraphics[width=\textwidth]{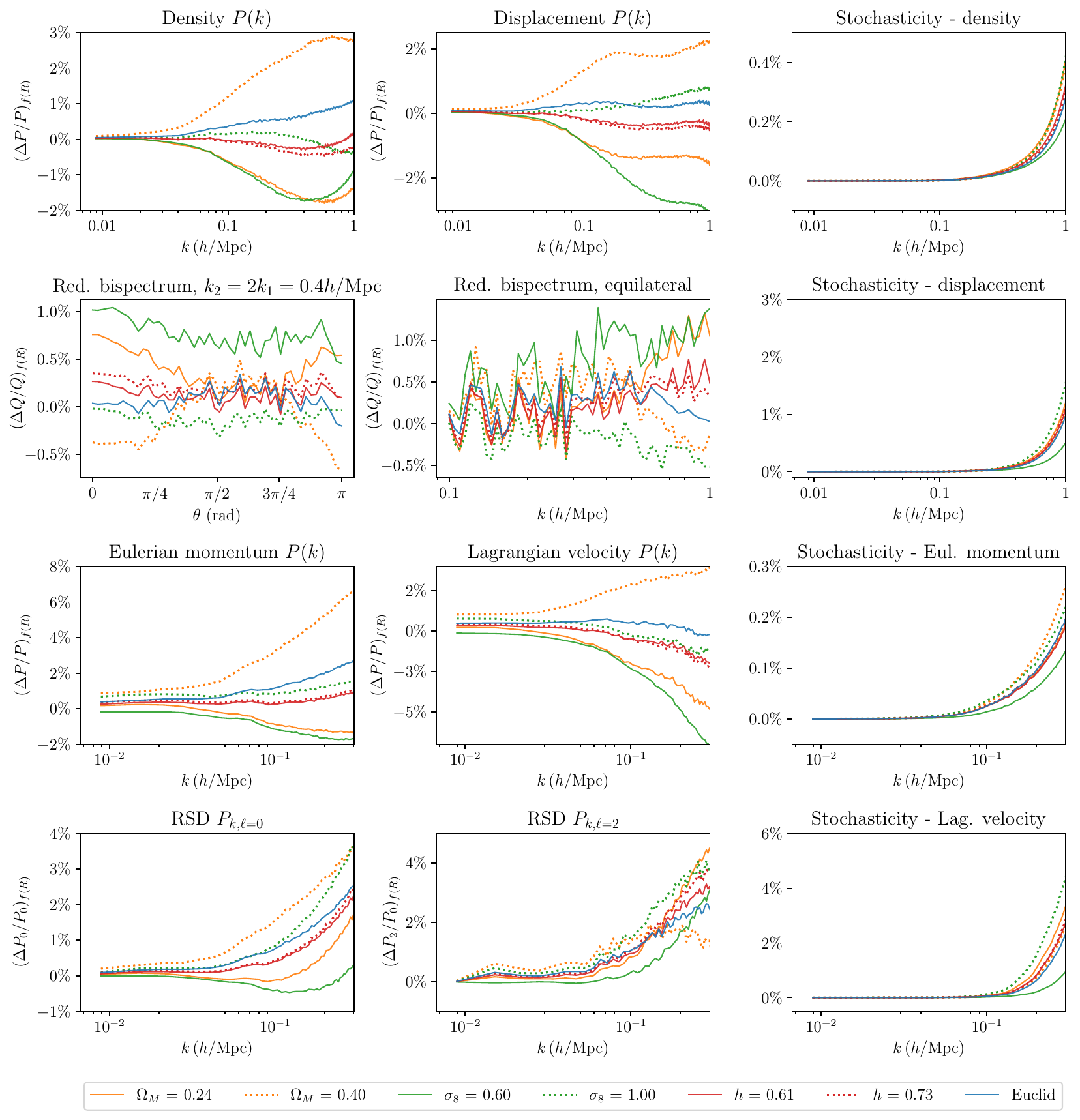}
	\caption{Performance of the emulator in extrapolations: the models used in this figure are defined in Table~\ref{tab:extrapolations}. We test the accuracy of the emulator on cosmological parameters it was not trained on, focusing on $\Omega_M$, $\sigma_8$ and $h$. We also test against the Euclid reference simulation \citep{EE2}. We find $\sim 3\%$ accuracy or better on the boost factor, and $\sim 1.5\%$ or better on the bispectrum, at $k < 1$ $h/$Mpc. In the RSD, we find $\sim 4\%$ or better in the monopole and quadrupole at $k < 0.3$ $h/$Mpc. Stochasticity is low in the Eulerian density and momentum fields -- less than $\sim 0.4\%$ at $k < 1$ $h/$Mpc and $k < 0.3$ $h/$Mpc. In the Lagrangian displacement and velocity, we find stochasticity under $\sim 2\%$ at $k < 1$ $h/$Mpc for the former, and under $\sim 4\%$ in the latter.}
	\label{fig:results_extrapolations}
\end{figure*}

\begin{table}
	\centering
	\caption{Cosmological models used in extrapolation. The first line (labelled ``standard'') shows the model used to produce the training, validation and test datasets. In subsequent lines, when parameters are not quoted, they are the same as in the ``standard'' line. We test the performance of our machine learning model when applied to models it was not trained on, each varying $\Omega_M$, $\sigma_8$ and $h$. Finally, we consider the Euclid reference cosmology \citep{EE2}, without massive neutrinos. For all the models in this table, \fRnot$=10^{-5}$.}
	\label{tab:extrapolations}
	\begin{tabular}{lccccc} 
		\hline
		label & $\Omega_M$ & $\sigma_8$ & $h$ & $\Omega_b$ & $n_s$ \\
		\hline
		standard                    & 0.281 & 0.842 & 0.697 & 0.046 & 0.971 \\
		$\Omega_M = 0.24$ & 0.24   &           &           &           &           \\
		$\Omega_M = 0.40$ & 0.40   &           &           &           &           \\
		$\sigma_8 = 0.60$    &           & 0.60   &           &           &           \\
		$\sigma_8 = 1.00$    &           & 1.00   &           &           &           \\
		$h = 0.61$                 &           &           &  0.61  &           &           \\
		$h = 0.73$                 &           &           &  0.73  &           &           \\
		Euclid                        & 0.319 & 0.816 & 0.67   & 0.049 & 0.96 \\
		\hline
	\end{tabular}
\end{table}

In this Section, we show the performance of our emulator when tested against cosmological parameters that it did not see in the training set, where only the style parameter \fRnot was varied, whilst the other parameters were kept fixed. 

The boost factor has a weak dependence on $\Omega_M$ and $\sigma_8$ \citep{Fitting_formulae_fR}, while showing relative insensitivity to the other parameters. For this reason, we focus on $\Omega_M$ and $\sigma_8$, choosing considerably different values: $\Omega_M \in \{ 0.24, 0.40 \}$ and $\sigma_8 \in \{ 0.60, 1  \}$. The range of $\Omega_M$ was chosen to match that of \texttt{EuclidEmulator2} \citep{EE2}.

Because of the observed discrepancy in the measured value for $H_0$ across different probes \citep{H0_1,H0_2,H0_3,H0_4}, we also to choose to vary $h$ in the range $h \in \{ 0.61, 0.73 \}$, to check that our emulator can be used in studies examining the origin of this tension. The chosen range for $h$ also matches that of \texttt{EuclidEmulator2}.

Finally, we focus on the Euclid reference cosmology $\{ \Omega_M=0.319, \sigma_8=0.816, h=0.67, \Omega_b=0.049, n_s=0.96 \}$ \citep{EE2}, without including massive neutrinos\footnote{Note that the boost is insensitive to massive neutrinos
\citep{Fitting_formulae_fR}.}. In this case, a total of five parameters are changed compared to the training set.

The seven models we test against are summarised in Table~\ref{tab:extrapolations}, whereas the results of this test on extrapolation are shown in Figure~\ref{fig:results_extrapolations}.

For all the models, we find $\sim 3\%$ accuracy or better on the boost factor, and $\sim 1.5\%$ or better on the bispectrum, at scales $k < 1$ $h/$Mpc. In the RSD, we find an agreement of $\sim 4\%$ or better in both the monopole and quadrupole at $k < 0.3$ $h/$Mpc. We see very low stochasticity in the Eulerian density and momentum field -- less than $\sim 0.4\%$ at $k < 1$ $h/$Mpc and $k < 0.3$ $h/$Mpc, respectively. In the Lagrangian displacement and velocity, we find stochasticity under $\sim 2\%$ at $k < 1$ $h/$Mpc for the former, and under $\sim 4\%$ in the latter.

Overall, we find very good agreement even in extrapolation.

\section{Conclusions} \label{sec:Conclusion}
Stage IV experiments will soon deliver exquisite data on the distribution of matter in the Universe. The potential to learn about the true nature of dark energy is enormous, but we cannot succeed unless the new data are supplemented by theoretical predictions of matching quality. Only the full output of N-body simulations currently raises to the task.

In this work, we have presented an emulator capable of turning the positions and velocities of a \LCDM\ N-body simulation into a more expensive one run under modified gravity. We assessed the performance of our emulator by computing several summary statistics from the predicted and target snapshots, finding $1\%$ accuracy in the nonlinear matter power spectrum up to scales of $k \sim 1$ $h/$Mpc. When compared against the emulator \texttt{eMantis}, based on the independent AMR code \texttt{ECOSMOG}, the predicted boost factor was still within 1.5\% accuracy at the same scales.

We additionally find $\sim 1\%$ accuracy in the bispectrum, which is sensitive to phases. The predicted RSD monopole and quadrupole from the emulated velocity field agree with the target ones within 4\% at scales $k < 0.3$  $h/$Mpc. Additionally, we find little stochasticity in both the emulated density and momentum fields.

Separately, we find considerable improvement in the predicted snapshots over rescaling techniques, based on several summary statistics. We also tested the emulator on cosmologies that it did not see during training and still found good agreements with target simulations. 

When used in tandem with \LCDM\ field-level emulators such as \mtm{}, our network is ideally positioned to bring transformative improvements in the accuracy of tests of gravity at cosmological scales. To this end, it can be readily incorporated into field-level inference methods like the Bayesian Origin Reconstruction from Galaxies (\texttt{BORG}) algorithm \citep{BORG}.

\section*{Acknowledgements}

DS thanks Yin Li, Drew Jamieson, Toby Maule, Baojiu Li for extremely helpful discussions. The authors thank the anonymous referee for helpful comments and suggestions.

DS, KK and XMA are supported by the STFC grant ST/W001225/1.

Numerical computations were carried out on the \texttt{Sciama} High Performance Computing (HPC) cluster, which is supported by the Institute of Cosmology and Gravitation, the South-East Physics Network (SEPNet) and the University of Portsmouth.

This paper is dedicated to the memory of Graziella Temporin.

For the purpose of open access, we have applied a Creative Commons Attribution (CC BY) licence to any Author Accepted Manuscript version arising. 

\section*{Data Availability}
Supporting research data and code are available upon reasonable request from the corresponding author.



\bibliographystyle{mnras}
\bibliography{references} 



\appendix

\section{Calibration of the COLA simulations} \label{appendix:eMantis}

In this Appendix, we give further details on the calibration of our \cola{} simulations; in particular, we discuss how we chose the values for the screening efficiencies reported in Table~\ref{tab:se}.

For every value of \fRnot, we calibrated the screening efficiency separately, sampling it at intervals of $0.1$. For every considered value of the screening efficiency, we ran two pair-fixed simulations for \LCDM\ and $f(R)$ gravity, possessing the same cosmological parameters and initialised with the same seed. We additionally set the starting amplitudes of all modes to be exactly those of the initial power spectrum, removing any spurious effects from cosmic variance. For every pair, we then computed the modified gravity boost $B_{k,\text{\texttt{FML}}} = P_{k,\text{\texttt{FML}}}^{\mathrm{(MG)}} / P_{k,\text{\texttt{FML}}}^{\mathrm{(\Lambda CDM)}}$, subsequently averaging it between the two pair-fixed simulations.

The averaged boost $ \langle B_{k,\text{\texttt{FML}}} \rangle$ was then compared against the \texttt{eMantis} prediction $B_{k,\text{\texttt{eMantis}}}$, determining the two loss functions:
\begin{equation} \label{eq:calibration_loss1}
	\ell_{\infty} = \max_{k} \left| B_{k,\text{\texttt{eMantis}}} - \langle B_{k,\text{\texttt{FML}}} \rangle \right|,
\end{equation}
 and
 \begin{equation} \label{eq:calibration_loss2}
 	\ell_2 = \sqrt{ \sum_k \left( B_{k,\text{\texttt{eMantis}}} - \langle B_{k,\text{\texttt{FML}}} \rangle \right)^2 },
 \end{equation}
 where the wavenumber $k$ varied in the range $0.1$ $h/$Mpc $ < k < 1$ $h/$Mpc. In these equations, the $\ell_{\infty}$ norm is the maximum absolute difference at any single $k$ -- and is a measure of local deviations from \texttt{eMantis} -- whereas the $\ell_2$ norm gives a more global measure of agreement.
 
 We then combined the losses in Eqns.~\eqref{eq:calibration_loss1}-\eqref{eq:calibration_loss2} into the quantity:
 \begin{equation} \label{eq:calibration_overall_loss}
 	\ell = \ell_2 \times \ell_{\infty},
 \end{equation}
to obtain a trade-off between these two measures of accuracy that reflected both global and local agreement with \texttt{eMantis}.
 
Using the loss in Eq.~\ref{eq:calibration_overall_loss}, we then performed a $\chi^2$ analysis on the sampled screening efficiencies, obtaining the values in Table~\ref{tab:se}.

\section{Tests on other values of \fRnot} \label{appendix:frac_fR0}

\begin{figure*}
	\centering
	\includegraphics[width=\textwidth]{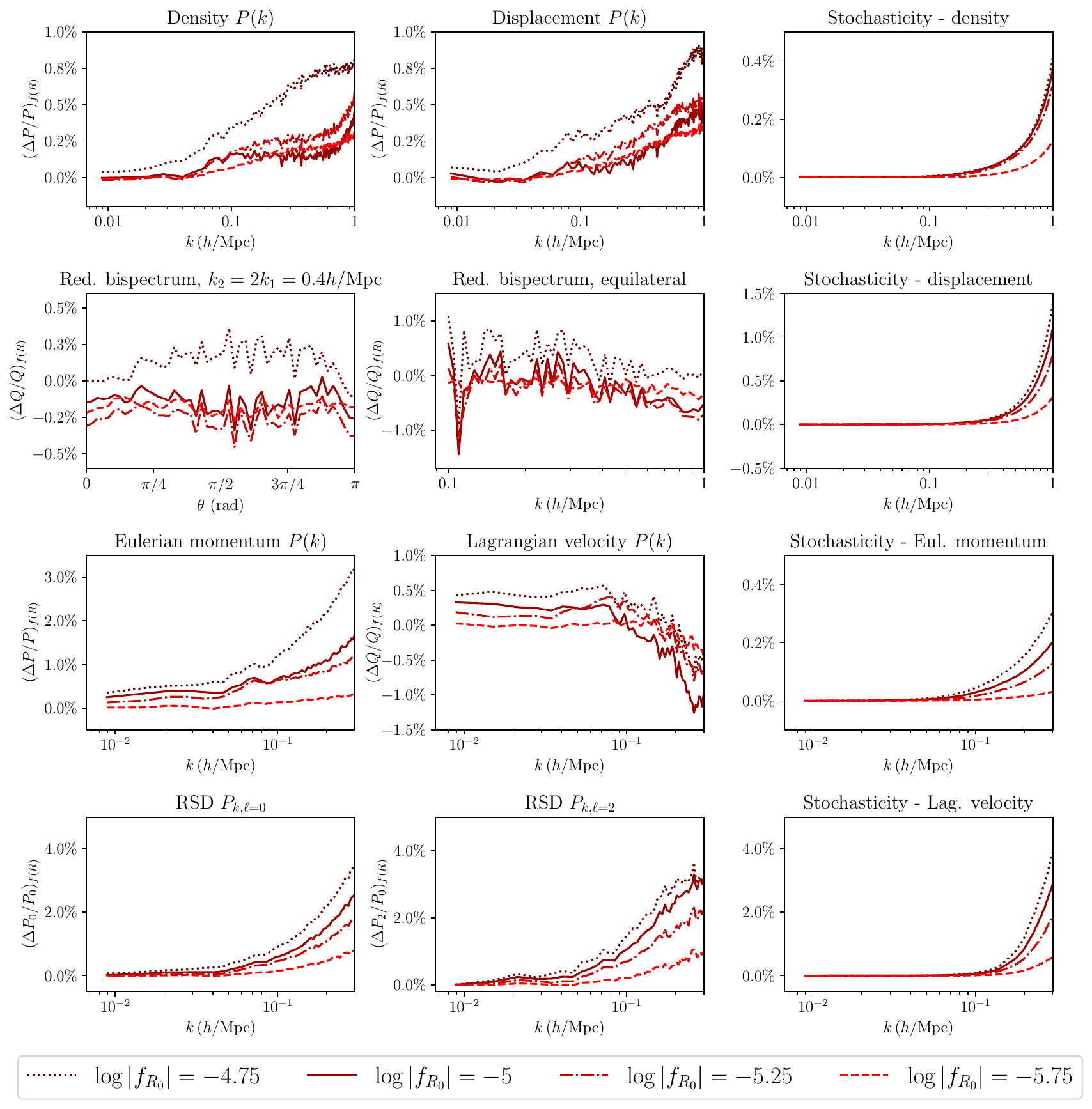}
	\caption{The emulator's performance on \fRnot values outside those employed throughout this work. The approximation of using a screening efficiency is not suited to resolve arbitrarily finely spaced values of \fRnot: consequently, we restrict ourselves to the values listed in Table~\ref{tab:se}, which are logarithmically spaced by $0.5$. In this plot, we show the result of testing our emulator on intermediate values: $\log{\left| f_{R_0} \right|} \in \{-4.75, -5.25, -5.75\}$ (dotted, dash-dotted and dashed lines), also showing $\log{\left| f_{R_0} \right|}=-5$ (solid line) for comparison. The accuracy of the emulator is comparable across all the \fRnot values displayed. Moreover, the error on the summary statistics depends monotonically on \fRnot, becoming larger as the mapping is more difficult to learn, indicating that the network is not overfitting the \fRnot dependence.}
	\label{fig:frac_fR0}
\end{figure*}

The screening efficiency \citep{Winther_2015,FML} used in the \fml\ library is a convenient tool to speed up modified gravity simulations, providing an excellent trade-off of speed and accuracy. However, it was not originally designed to characterise arbitrarily finely spaced values of the modified gravity strength \fRnot.

For this reason, we train and validate our emulator using the \fRnot values reported in Table~\ref{tab:se}, which are logarithmically spaced by $0.5$. All the results shown in Sec.~\ref{sec:Results} also use these values. In this Appendix, we show that our emulator achieves the same level of accuracy on intermediate values, indicating its ability to interpolate effectively between the discrete training points.

In Figure~\ref{fig:frac_fR0}, we test the emulator on $\log{\left| f_{R_0} \right|} \in \{-4.75, -5.25, -5.75\}$ (dotted, dash-dotted and dashed lines, respectively), while also showing the $\log{\left| f_{R_0} \right|}=-5$ result in a solid line for reference (this value is included in all the figures of Sec.~\ref{sec:Results}). The screening efficiency used for these new values of \fRnot, determined as detailed in Appendix~\ref{appendix:eMantis}, is $\{2.2, 2.2, 2.0\}$, respectively (the screening efficiencies used for \fRnot$=-4.75,-5.25$ are the same within the significant figures we use).

We compare emulated and target simulations against all the summary statistics previously included -- i.e. the power spectrum and stochasticities of the density, displacement, Eulerian momentum and Lagrangian velocity fields, two configurations of the reduced bispectrum (equilateral and $k_2=2 k_1 = 0.4$ $h/$Mpc), and the monopole and quadrupole of the redshift-space distortions.

The accuracy of the emulator is comparable to that shown in Sec.~\ref{sec:Summary_statistics}. Moreover, Figure~\ref{fig:frac_fR0} shows that the error on the summary statistics depends monotonically on \fRnot, becoming larger as the mapping is more difficult to learn, indicating that the network is not overfitting the \fRnot dependence.


\bsp	
\label{lastpage}
\end{document}